\documentclass[twocolumn,showpacs,preprintnumbers,amsmath,amssymb]{revtex4}
\usepackage{graphicx}
\usepackage{dcolumn}
\usepackage{bm}
\usepackage{mathrsfs}
\usepackage{amsmath,amssymb,graphics}

\begin{document}

 \title{Anomaly analysis of Hawking radiation from 2+1 dimensional
  spinning black hole}

\author{Shao-Wen Wei\footnote{E-mail: weishaow06@lzu.cn} ,
        Ran Li\footnote{E-mail: liran05@lzu.cn},
        Yu-Xiao Liu\footnote{Corresponding author. E-mail: liuyx@lzu.edu.cn}
        and Ji-Rong Ren}
\affiliation{
    Institute of Theoretical Physics, Lanzhou University,
           Lanzhou 730000, P. R. China}

\begin{abstract}
Considering gravitational and gauge anomalies at the horizon, a
new successful method that to derive Hawking radiations from black
holes has been developed recently by Wilczek et al.. By using the
dimensional reduction technique, we apply this method to a
non-vacuum solution, the 2+1 dimensional spinning black hole. The
Hawking temperature and angular velocity on the horizon are
obtained. The results may partially imply that this method is
independent of the gravity theory, the dimension of spacetime and
the topological structure of the event horizon.

 \end{abstract}


 \maketitle

\section{introduction}

The discovery of Hawking radiation \cite{hawking} reveals that the
black hole is not completely black but can emit radiation from its
event horizon like a black body at the temperature
$T=\frac{\kappa}{2\pi}$. This raises the interest of investigation
for the Hawking radiation. Apart from the original derivation, one
of the derivations is the tunneling method, which based on pair
creations of particles and antiparticles near the horizon through
calculating WKB amplitudes for classically forbidden trajectories
\cite{Parikh:1999mf,tunneling}. Some new important results of
tunneling method can be found in
\cite{Banerjee20093plb,Akhmedova2009plb}.

A couple of years ago, a new method to obtain the Hawking radiation
was proposed by Robinson and Wilczek et al \cite{Robinson2005prl}.
The facts, a black hole's horizon is a one way membrane and the
effective theory near it is a two-dimensional theory, lead to the
gauge and gravitational anomalies for the currents near a black
hole's horizon. By solving these anomaly equations with the regular
conditions, one may regard the Hawking radiation as the compensation
of the anomaly to break the classical symmetry. The result was first
obtained from a static and spherically symmetric Schwarzschild-type
black hole. Soon, this method is extended to charged and rotating
black holes \cite{Iso2006prl,Iso2006prd}. In fact, there are two
anomalous currents, the consistent current and the covariant
current. The consistent current anomaly equation satisfies the Wess
Zumino consistency condition. But under a gauge transformation, it
does not transform covariantly. On the other hand, Banerjee and
Kulkarni \cite{Banerjee2008prd,Banerjee2008plb} introduced another
covariant current, which transforms under gauge transformation, but
does not satisfy the Wess Zumino consistency condition. As was shown
later, the anomaly cancellation method has been successfully
extended to other black hole cases
\cite{Vagenas2006jhep}-\cite{Iso2008ijmpa}. For these black holes,
with the different gravity theory, the dimension of spacetime, the
topological structure of the event horizon, the result is universal.
It also can be seen that the thermal distribution of Hawking
radiation can be completely reproduced by investigating
transformation properties of the higher-spin currents under
conformal and gauge transformations
\cite{Iso2007prd,Iso2008ijmpa,Bonora2008jhep}.

In this letter, we would like to investigate the spinning black hole
using this method. Different from the generally black holes
solutions, the metric component $g_{tt}$ has no singular when
$r\rightarrow r_{h}$ (the radius of event horizon). This non-vacuum
solution was presented by Chan and Mann in
\cite{Chan1996plb2,Chan1996plb}. Other spinning Solutions in 2 + 1
dimensional Gravity can be found in \cite{Chen1999npb}. Its entropy
was also studied in \cite{Lee1996plb}, where they showed that the
entropy diverges when $h\rightarrow 0$, where $h$ is the radial
coordinate distance from the horizon to the brick wall. Recently, it
is found that the wave equations of a massless scalar field is shown
to be exactly solvable in terms of hypergeometric functions for a
slowly spinning black hole and the asymptotic form of the
quasinormal frequencies are obtained \cite{Fernando2009}. Base on
the above considerations, we want to known whether the Hawking
radiation can be obtained from the anomaly method for the spinning
black hole under the gravity theory. Firstly, we will perform a
dimensional reduction of an action given by a complex scalar field
minimally coupled to gravity in the background of a
(2+1)-dimensional spinning black hole. The result shows that the
theory is reduced to an effective theory of an infinite collection
of (1+1)-dimensional scalar fields near the horizon. We found that,
after the dimensional reduction, an effective $U(1)$ gauge field is
generated by an angular isometry. The azimuthal quantum number $m$
serves as the charge of each partial wave, and this result accords
with that of \cite{Iso2006prd}. Through studying the gauge and
gravitational anomalies, we obtain the Hawking flux.

The paper is organized as follows. In section \ref{nonKK}, we
review the basic properties of (2+1)-dimensional spinning black
hole and carry out the dimensional reduction. It's Hawking
radiation is derived via anomalies in section \ref{Hawking}.
Finally, the paper ends with a brief summary.

\section{Quantum field in spinning black hole}
\label{nonKK}

In this section we will give a brief review of the spinning black
hole and carry out the dimensional reduction. The action which leads
to spinning black hole solutions is \cite{Chan1996plb}
\begin{equation}
S=\int d^{3}x\sqrt{-g}\bigg(\mathcal{R}
 -4(\nabla \phi)^{2}
 +2e^{b\phi}\Lambda\bigg),
\end{equation}
where $\Lambda$ is the cosmological constant and $\phi$ is the
dilaton field.

A family of spinning black hole solutions were presented in
\cite{Chan1996plb} with the metric given by
\begin{eqnarray}
ds^2
  =&& -\bigg(\frac{8 \Lambda r^N}{(3 N -2)N}+
   \eta r^{1-\frac{N}{2}}\bigg) dt^2\nonumber\\
   &&+\frac{1}{\bigg[\frac{8\Lambda r^N}{(3N-2)N}
   +\bigg(\eta-\frac{2\Lambda \gamma^2}{(3N-2)N\eta} \bigg)
   r^{1-\frac{N}{2}}\bigg]}dr^2 \nonumber\\
  && - \gamma r^{1-\frac{N}{2}}dt d\theta +\bigg(r^N -
   \frac{\gamma^2}{4\eta} r^{1-\frac{N}{2}}\bigg)d\theta^2,
\end{eqnarray}
where the mass $M$, angular momentum $J$ and parameter $\eta$ are
given by
\begin{eqnarray}
M &=&\frac{N}{2}\bigg[\frac{2\Lambda \gamma^2}{(3N-2)N \eta}
  \bigg(\frac{4}{N}-3 \bigg)-\eta \bigg], \\
J &=& \frac{3N-2}{4}\gamma,\\
\eta&=&-\frac{M}{N}-\sqrt{\frac{M^2}{N^2}+\left(\frac{4}{N}-3\right)
  \frac{2\Lambda \gamma^2}{(3N-2)N}}.
\end{eqnarray}
Here, $\eta$ and $\gamma$ are integration constants. For the action
contains a static ¡°dilaton fluid¡± whose energy momentum tensor is
nowhere vanishing, it is a non-vacuum solution. For the dilaton
coupling parameters $N=1$, the quasi-normal modes and the area
spectrum have been studied in detail in \cite{Fernando2009}. In this
paper, we will also focu on the case of $N$=1 and the metric for
this spinning black hole is reduced to
\begin{equation}
 ds^2= - f(r)  dt^2 +  \frac{dr^2}{ h(r) } - 4 J r d \theta dt +
     p^2(r) d \theta^2,
\end{equation}
with
\begin{eqnarray}
 &&f(r) =\left(8 \Lambda r^{2}-( M + \sqrt{ M^2 + 32 \Lambda J^2})r\right),\\
 &&h(r) =\frac{4 \Lambda r- M}{2 r },\\
 &&p^{2}(r)=\left(r^{2} + \frac{(-M +\sqrt{M^{2}+32\Lambda J^{2}})}{8 \Lambda}
    r\right).
\end{eqnarray}
After some calculations, we found there exists a special
relationship between these metric functions,
\begin{equation}
 f p^{2}+4J^{2} r^{2}=4 r^{4} h. \label{simplify}
\end{equation}
The horizon is located at $h(r)=0$ and is given by
\begin{equation}
 r_{H}=\frac{M}{4 \Lambda}.
\end{equation}
It is clear that $f(r)$ and $p(r)$ are non-singular at $r=r_{H}$.
The determinant of the metric is
\begin{equation}
 g=-4 r^{4},
\end{equation}
and non-zero metric coefficients can be simplified with
(\ref{simplify}),
\begin{eqnarray}
 &&g^{00}=-\frac{p^{2}}{4 h r^{4}},\nonumber\\
 &&g^{02}=g^{20}=-\frac{J}{2 h r^{3}},\nonumber\\
 &&g^{11}=h, \label{spin metric}\\
 &&g^{22}=\frac{f}{4 h r^{4}}.\nonumber
\end{eqnarray}
Next, we will carry out the dimensional reduction for this metric
background. Considering a action functional constructed from a
complex scalar field:
\begin{equation}
 S=\frac{1}{2} \int d^{3}x \sqrt{-g}\Phi^* \nabla^{2}\Phi.
 \label{action}
\end{equation}
The Laplace-Beltrami operator $\nabla^{2}$ is defined by
\begin{equation}
 \nabla^{2}\equiv
 \frac{1}{\sqrt{-g}}\partial_{\mu}(\sqrt{-g}g^{\mu\nu}\partial_{\nu}).
 \label{operator}
\end{equation}
Substituting Eqs. (\ref{spin metric}) and (\ref{operator}) into
(\ref{action}), we can obtain
\begin{eqnarray}
 S=\frac{1}{2} \int d^{3}x \Phi^* \bigg[&-&\frac{p^{2}}{2 h r^{2}} \partial_{t}^{2}
   -\frac{2J}{h r}\partial_{t} \partial_{\theta}\nonumber\\
   &+&\partial_{r}(2 r^{2} h
   \partial_{r})+\frac{f}{2hr^{2}} \partial_{\theta}^{2}\bigg] \Phi.
\end{eqnarray}
Performing the partial wave decomposition,
$\Phi=\sum_{m}\varphi_{m}(r,t)e^{im \theta}$ and integrating the
angle coordinate part, the above action will be reduced to
\begin{eqnarray}
 S=\pi \sum_{m}\int dt dr \varphi_{m}^*
   \bigg[&-&\frac{p^{2}}{2 h r^{2}}
          (\partial_{t}+\frac{2imJr}{p^{2}})^{2}\nonumber\\
         &+&\partial_{r}(2 r^{2} h \partial_{r})-m^{2}\frac{2 r^{2}}{p^{2}} \bigg] \varphi_{m}.
  ~~~~ \label{2 action}
\end{eqnarray}
It can be seen that the action is reduced to an infinite set of the
scalar fields $\varphi_{m}$ on a two-dimensional space. It is
convenient to define a tortoise coordinate to discuss the behavior
of the action near the horizon. The tortoise coordinate can be
defined as
\begin{equation}
 \frac{d r_{*}}{dr}=\frac{1}{h}.
\end{equation}
After this coordinate transform, the last term appears in (\ref{2
action}) is $-\frac{2m^{2} r^{2}h}{p^{2}}$, which will vanish as
$r\rightarrow r_{H}$. Ignoring it, the final effective
two-dimensional action can be rewritten as
\begin{eqnarray}
 S=\pi \sum_{m}\int dt dr \varphi_{m}^* p
     \bigg[&-&\frac{p}{2 h r^{2}}
             (\partial_{t}+\frac{2imJr}{p^{2}})^{2}\nonumber\\
           &+&\partial_{r}(\frac{2 r^{2} h}{p} \partial_{r}) \bigg]\varphi_{m}.
\end{eqnarray}
After undergoing the dimensional reduction near the horizon, we
could see that each partial wave of the scalar field can be
effectively described by an infinite collection of complex scalar
field in the background of a (1+1)-dimensional metric:
\begin{eqnarray}
 ds^{2}=-F(r) dt^{2}+F^{-1}(r)dr^{2}, \label{2metric}
\end{eqnarray}
where $F=\frac{2 hr^{2}}{p(r)}$. The dilaton field and gauge
potential $\mathcal{A}_{\mu}$ is given by
\begin{eqnarray}
 \Psi=2\pi p, \mathcal{A}_{t}=-\frac{2Jr}{p^{2}},\mathcal{A}_{r}=0.
\end{eqnarray}

\section{Hawking radiation of spinning black hole}
\label{Hawking}

For the $U(1)$ gauge current $\mathcal{J_\mu}$, the consistent form
of anomaly is given by
\begin{eqnarray}
 \nabla_\mu \mathcal{J^\mu} =\pm
 \frac{m^2}{4\pi}\epsilon^{\mu\nu}\partial_\mu \mathcal{A}_\nu,\label{Ju}
\end{eqnarray}
where $+(-)$ corresponds to the left(right)-handed fields and
$\epsilon^{\mu\nu}$ is antisymmetric with $\epsilon^{01}=1$. The
current $\mathcal{J^\mu}$ appeared in (\ref{Ju}) is not a covariant
current. However, covariant current can be defined as
\begin{eqnarray}
 \tilde{J}^\mu =\mathcal{J^\mu} \mp
 \frac{m^2}{4\pi}\mathcal{A}_\lambda\epsilon^{\lambda\mu},
\end{eqnarray}
which satisfies
\begin{eqnarray}
 \nabla_\mu\tilde{J}^\mu=\pm\frac{m^2}{4\pi}\epsilon_{\mu\nu}\mathcal{F}^{\mu\nu}.
\end{eqnarray}
The consistent current can be written as a sum of two regions,
outside the horizon and near the horizon ,
\begin{eqnarray}
 \mathcal{J^\mu}=\mathcal{J^\mu_{(O)}}\Theta_+(r)+\mathcal{J^\mu}_{(\mathcal{H})}H(r),
\end{eqnarray}
where $\Theta_+(r)=\Theta(r-r_H-\epsilon)$ and $H(r)=1-\Theta_+(r)$
are step functions which are defined in the region $r \geq r_{H}$.
$\mathcal{J^\mu_{(O)}}(r)$, the current outside the horizon, is
conserved and satisfies
\begin{eqnarray}
 \partial_r \big[\mathcal{J}^r_{\mathcal{(O)}}\big]=0. \label{eq1}
\end{eqnarray}
While $\mathcal{J^\mu}_{(\mathcal{H})}(r)$, the current near the
horizon, satisfies the anomalous equation
\begin{eqnarray}
 \partial_r \big[  \mathcal{J}^r_{(\mathcal{H})}\big]=\frac{m^2}{4\pi}\partial_r \mathcal{A}_t. \label{eq2}
\end{eqnarray}
Eqs. (\ref{eq1}) and (\ref{eq2}) can be easily integrated as,
respectively,
\begin{eqnarray}
 \mathcal{J}^r_{\mathcal{(O)}}&=&c_{\mathcal{O}},\nonumber\\
 \mathcal{J}^r_{(\mathcal{H})}&=&c_{\mathcal{H}}+\frac{m^2}{4\pi}(\mathcal{A}_t(r)-\mathcal{A}_t(r_H)),
\end{eqnarray}
where $c_\mathcal{O}$ and $c_\mathcal{H}$ are integration constants.
$c_\mathcal{O}$ is the value of the electric flux.

Considering a variation of quantum effective action $W$ under a
gauge transformation with gauge parameter $\zeta$,
\begin{eqnarray}
 -\delta W&=&\int d^2x \zeta\nabla_\mu \mathcal{J}^\mu\nonumber\\
 &=&\int d^2x \zeta \bigg[\partial_r \big(\frac{m^2}{4\pi}\mathcal{A}_t
     H(r)\big)\nonumber\\
 & &+\delta(r-r_H-\epsilon)\big(
 (\mathcal{J}^r_{\mathcal{(O)}}-\mathcal{J}^r_{(\mathcal{H})})+\frac{m^2}{4\pi}\mathcal{A}_t\big)\bigg],
 ~~~~
\end{eqnarray}
The coefficient of the delta function in the above equation should
vanish. This leads to the result
\begin{eqnarray}
 c_\mathcal{O}=c_\mathcal{H}-\frac{m^2}{4\pi}\mathcal{A}_t(r_H).
\end{eqnarray}
Imposing the condition that the covariant current vanishes at the
horizon, the value of the current can be fixed at the horizon
\begin{eqnarray}
 c_\mathcal{H}=-\frac{m^2}{4\pi}\mathcal{A}_t(r_H).
\end{eqnarray}
Then the electric flux is obtained
\begin{eqnarray}
 c_\mathcal{O}=-\frac{m^2}{2\pi}A_t(r_H)=\frac{m^2}{2\pi} \frac{16 J \Lambda}{(M+\sqrt{M^{2}+32\Lambda J^{2}})},
\end{eqnarray}
This agrees with the current flow associated with the Hawking
thermal (blackbody) radiation.

For the energy-momentum tensor, the anomalous Ward identity is given
by
\begin{eqnarray}
 \nabla_{\mu}\mathcal{T}^{\mu}_{\nu}=\mathcal{F}_{\mu\nu}\mathcal{J}^{\nu}+\mathcal{A}_{\nu}
   =\mathcal{F}_{\mu\nu}\mathcal{J}^{\nu}-\frac{1}{96\pi}\epsilon_{\mu\nu}\partial^{\mu}R
\end{eqnarray}
with $\mathcal{A}_{\mu}$ is the covariant gravitational anomaly. For
the metric (\ref{2metric}), the covariant anomaly is purely
time-like for $\mathcal{A}_{r}=0$. The energy-momentum tensor can be
decomposed as
\begin{eqnarray}
 \mathcal{T}^\mu_\nu=\mathcal{T}^\mu_{\nu\mathcal{(O)}}\Theta_+(r)+\mathcal{T}^\mu_{\nu(\mathcal{H})}H(r).
\end{eqnarray}
Outside the horizon, we get
\begin{eqnarray}
 \partial_r \big[\mathcal{T}^r_{t\mathcal{(O)}}\big]=\mathcal{J}^r_{\mathcal{(O)}}\partial_r
 \mathcal{A}_t.
\end{eqnarray}
Near the horizon, we have the anomalous equation
\begin{eqnarray}
 \partial_r \big[ \mathcal{T}^r_{t(\mathcal{H})}\big]=\big[\mathcal{J}^r_{(\mathcal{H})}\partial_r
 \mathcal{A}_t +\mathcal{A}_t\partial_r \mathcal{J}^r_{(\mathcal{H})}\big]+\partial_r
 \mathcal{N}^r_t,
\end{eqnarray}
where $\mathcal{N}^r_t$ is given by
\begin{eqnarray}
 \mathcal{N}^r_t=\frac{1}{96\pi}\epsilon^{\mu\nu}\partial_\nu\Gamma^r_{t\mu}.
 \label{Nrhotau}
\end{eqnarray}
Integrating these two equations, we obtain
\begin{eqnarray}
 \mathcal{T}^r_{t\mathcal{(O)}}&=&a_\mathcal{O}+c_\mathcal{O} \mathcal{A}_t,\nonumber\\
 \mathcal{T}^r_{t(H)}&=&a_\mathcal{H}+\int_{r_h}^r dr
 \partial_r\bigg(c_\mathcal{O}
 \mathcal{A}_t+\frac{m^2}{4\pi}\mathcal{A}_t^2+\mathcal{N}^r_t\bigg).
\end{eqnarray}

Under the infinitesimal general coordinate transformations, the
effective action transforms as
\begin{eqnarray}
 -\delta W&=&\int d^2x \xi^\tau\nabla_\mu \mathcal{T}^\mu_t\nonumber\\
 &=&\int d^2x \xi^t \bigg\{ c_\mathcal{O}\partial_r A_t
 +\partial_r\bigg[\big(\frac{m^2}{4\pi}\mathcal{A}_t^2+\mathcal{N}^r_t\big)H\bigg]\nonumber\\
 &&+\delta(r-r_H-\epsilon)\bigg[(\mathcal{T}^r_{t\mathcal{(O)}}-\mathcal{T}^r_{t(\mathcal{H})})
 +\mathcal{N}^r_t+\frac{m^2}{4\pi}\mathcal{A}_t^2\bigg]
 \bigg\}. \nonumber \\
\end{eqnarray}
The coefficient of the delta function term need to vanish at the
horizon,
\begin{eqnarray}
 a_\mathcal{O}=a_\mathcal{H}+\frac{m^2}{4\pi}\mathcal{A}_t^2(r_H)-\mathcal{N}^r_t(r_H).
\end{eqnarray}
Imposing a vanishing condition for the covariant energy-momentum
tensor at the horizon, which gives the equation
\begin{eqnarray}
 a_\mathcal{H}=2\mathcal{N}^r_t(r_H).
\end{eqnarray}
The total flux of the energy-momentum tensor is given by
\begin{eqnarray}
 a_\mathcal{O}=\frac{m^2}{4\pi}\mathcal{A}_t^2(r_H)+\mathcal{N}^r_t(r_H).
\end{eqnarray}
From (\ref{Nrhotau}) and metric (\ref{2metric}), we can calculate
\begin{eqnarray}
 \mathcal{N}^r_t(r_H)=\frac{1}{192\pi}\big(F\big)'^2\big|_{r_H}
 =\frac{\pi}{12\beta^2}.
\end{eqnarray}
So we have
\begin{eqnarray}
 a_\mathcal{O}=\frac{m^2}{4\pi}V^2+\frac{\pi}{12\beta^2},
\end{eqnarray}
where $V^2 = (\frac{2Jr}{p^{2}}\mid_{r=r_{H}})^{2}$. Note that the
flux is proportional to $(T_H)^2$ with $T_H =\frac{1}{\beta}$, the
Hawking temperature of the black hole. Its temperature and angular
velocity at the horizon are given by, respectively,
\begin{eqnarray}
 T_{H}&=&\frac{1}{4\pi}\bigg(\frac{1}{16\Lambda^{2}}
         +\frac{J^{2}}{M\Lambda(M+\sqrt{M^{2}+32\Lambda
          J^{2}})}\bigg)^{-\frac{1}{2}},\\
 \Omega&=&\mid \mathcal{A}_{t}\mid_{r=r_{H}}
   =\frac{2Jr_{H}}{p(r_{H})^{2}}=\frac{2J}{(r_{H}
   + \frac{(-M +\sqrt{M^{2}+32\Lambda J^{2}})}{8
   \Lambda})},\nonumber \\
\end{eqnarray}
the same with that of \cite{Fernando2009}. Until now, we have
calculated the gauge current and energy-momentum tensor flux. The
results show that these fluxes are exactly equivalent to Hawking
radiation from the event horizon.

\section{Summary}

In this paper, we obtain the Hawking flux from a non-vacuum
solution, (2+1)-dimensional spinning black hole, by using the method
of quantum anomalies. Firstly, by integrating the action given by a
scalar field minimally coupled to gravity in the background of black
hole near the horizon, we obtain a (1+1)-dimensional effective
action, which means physics near the horizon can be described with
an infinite collection of massless (1+1)-dimensional scalar fields.
It is also found that, after the dimensional reduction, an effective
$U(1)$ gauge field is generated by an angular isometry. The
azimuthal quantum number $m$ serves as the charge of each partial
wave. Then, considering the quantum anomalies near the horizon, we
obtain the exact Hawking flux. The results may partially imply that
this method is independent of the gravity theory, the dimension of
spacetime and the topological structure of the event horizon.

\section*{Acknowledgement}

This work was supported by Program for New Century Excellent
Talents in University, the National Natural Science Foundation of
China(NSFC)(No. 10705013), the Doctoral Program Foundation of
Institutions of Higher Education of China (No. 20070730055), the
Key Project of Chinese Ministry of Education (No. 109153) and the
Fundamental Research Fund for Physics and Mathematics of Lanzhou
University (No. Lzu07002).

\end{document}